\documentclass[preprint2]{aastex}

\slugcomment{To appear in Astronomical Journal}

\shorttitle{Optical variability of PKS 2005-489 and PKS 2155-304}
\shortauthors{Dominici et al.}

\begin{document}

\title{Long-term optical variability of the blazars PKS 2005-489 and
 PKS 2155-304}

\author{T.P. Dominici, Z. Abraham, R. Teixeira and P. Benevides-Soares}
\affil{Instituto de Astronomia, Geof\'\i sica e Ci\^encias 
Atmosf\'ericas, Universidade de S\~ao Paulo\\
Rua do Mat\~ao 1226, Cidade Universit\'aria, 05508-900,
 S\~ao Paulo, SP, Brazil\\}

\begin{abstract}

We present optical light curves for the period
1996-2000, of two of the brightest known EGRET BL Lac objects:  PKS 2005-489 and 
PKS 2155-304, 
the latter also one of the few known TeV sources. 
For both objects, quiescent epochs of
low level of variability were followed by active periods, without 
any indication of periodicity.
In PKS 2005-489, several variability events with duration of about 20 days were 
observed.
In PKS 2155-304 fast drops  and subsequent rises in luminosity occurred in time 
scales of days.
We proposed an explanation in which a region moving along
the relativistic jet is eclipsed by broad line region clouds or star clusters in 
the host galaxy. 
We compare our light curves with contemporaneous X-ray observations from All-Sky 
Monitor/RXTE. 
Correlations between optical and X-ray activity were not found in any of the 
sources at long time scales.
However in PKS 2005-489 possible correlations were observed in 1997 and 1998 at 
short time scales, 
with optical variability preceding X-rays
by 30 days in 1997 and succeeding them by about 10 days in 1998.  
 The analysis 
of the  SED, using the optical data presented here and
BeppoSAX contemporaneous observations obtained from the literature, shows only 
small shifts in the
synchrotron peak as the
 X-ray flux density changes. 
 \end{abstract}




\section{Introduction}

Variability studies are a basic tool to understand the
physical processes occurring in AGNs, especially those related to short time 
scales and
high amplitudes, as observed in BL Lacs.  
In this paper we made use of the five-year database of bright extragalactic
sources obtained with the meridian circle of the Abrah\~ao de Moraes Observatory 
(Valinhos, Brazil), between 1996 and 2000, to study light
variations of two EGRET BL Lac sources:
PKS 2005-489 and PKS 2155-304.

Both sources are high-frequency peaked BL Lacs (HBLs), with the synchrotron peak 
located
between UV and X-ray wavelengths and were also detected by EGRET in the 100 MeV 
range
(Lin et al. 1999). Moreover,  PKS 2155-304 is the only southern 
hemisphere BL Lac object detected  at TeV energies (Chadwick et al. 1999), while 
PKS 2005-489, also a strong candidate for detection at this energy range with the 
available
 instrumentation (Stecker et al. 1996), has not yet been detected (Chadwick et al. 
2000).

The compact nature of PKS 2005-489 at radio frequencies was confirmed by VLBI 
observations at 
5 GHz  (Shen et al. 1998).
Piner \& Edwards (2004) were able to resolve PKS 2155-304 into a core and a jet 
component 
 at 15 GHz and obtained an upper limit of 4$c$ for its apparent velocity.  
 Since even lower velocities were found for the other two TeV blazars observed in 
that work, 
the authors
concluded that stationary or subluminal velocities seem to be a characteristic of 
this kind of sources, 
in contrast 
with the majority of blazars, for which parsec scale jet components are detected. 

The interest in PKS 2005-489 increased after the detection of two strong X-ray 
flares in 1998 by the
Rossi X-Ray Timing Explorer (RXTE, Remillard, 1998, Perlman et al. 1998), however
few optical data are available for this object. The only published information 
about  
long term variations was presented by Wall et al. (1986), who found a change of 
0.5 magnitudes in the B 
band in two sets of observations spaced by one year. In time scales of a few days, 
observations in 1994 by 
Heidt \& Wagner (1996) showed
 10\% amplitude variations in the R band. Similar results were obtained by 
Rector \& Perlman (2003) in observations taken some days after a third X-ray flare 
in September 2000, while 
microvariability (variability of minutes to hours) was not detected neither at 
that epoch nor in 1997 
(Romero et al. 1999). 

More observations at several wavelengths are available for PKS 2155-304. 
Using the optical and near infrared data from the literature after 1970, Fan \& 
Lin (2000)
 suggested the existence of 4 and 7 year periodicity in the light curves. 
 Variability at optical wavelengths in time scales of
less than 15 minutes during 1990 was reported by Heidt, Wagner \& Wilhein-Erkes 
(1997) and during 1995 
by Paltani et al. (1997). Brightness variations in time scales from few days to 
minutes 
were also observed in two large multiband monitoring campaigns, from radio to 
X-rays,  
during the last years (e.g. Edelson et al. 1995, 
Pesce et al. 1997, Urry et al. 1997)  and the existence of lags between 
variability at X-rays
and ultraviolet wavelengths verified.
However, no microvariability was found by Heidt et al. (1997) in 1994 and by 
Romero et al. (1999) in
1997 and 1998, showing the existence
of a duty cycle for this kind of variability.

In this paper we show that both objects present long  time scale variability, 
alternating between
high and low activity. 
In Section 2 we describe the main aspects of
the instrumentation used in this work and the data reduction procedures. The 
resulting 
differential light curves are shown in Section 3 and an analysis of the  
correlation
between optical and All-Sky Monitor X-ray data is presented in Section 4. In 
Section 5 
we propose a scenario to explain  observed dips in the 1999 light curve of PKS 
2155-304. 
The implication of the data in the spectral energy distribution of the sources is 
discussed
 in Section 6. Finally, in Section 7 we summarize our results.

\section{Observations and data reduction}

The observations were made with the Askania-Zeiss meridian circle, 
installed at the Abrah\~ao de Moraes Observatory, Valinhos, Brazil ($\phi$ = 
$46^{o}$ 58' 03'', 
$\lambda$ = $-23^{o}$ 00' 06''). The instrument  
is a 0.19 m refractor with a focal 
distance of 2.6 m and has a CCD Thomson 7895A detector
($512 \; \times \, 512$ pixel, with a  scale of $1.5"/$pixel). Observations are
 made in drift scan mode, where the speed of charge transfer by the CCD is the 
same
as the speed of stellar transit for a fixed instrument position. In this case, the 
integration
 time for a declination $\delta$ is given by
$ t_{\rm int} = 51  \sec \delta $ seconds (Viateau et al. 1999, Dominici et al. 
1999).   
The observed field has a width of $13'$ in declination and an arbitrary 
value in right ascension. 

The filter used in the instrument is wider than the 
standard Johnson filter V, with a better coverage for longer wavelengths.
The transformation between
the magnitudes at the different filters can be estimated from (Dominici et al. 
1999):

$$ V - V_{\rm Val} = 0.14 - 0.07(B-V) $$

\noindent
where $V_{\rm Val}$ designates the magnitudes obtained
 with the Valinhos system and $V$ and $B$ are the magnitudes in the standard
 Johnson system.

Our observations started in 1996 and extended to 2000 for PKS 2005-489 and
to 1999 for PKS 2155-304. 
The data are part of an astrometric program directed
 to the determination 
of a reference system based on extragalactic sources (Teixeira et al. 1998).
Table 1 presents basic information about the observed objects and the 
observational program.  
The positions, V magnitudes and redshifts 
were taken from V\'eron-Cetty \& V\'eron (1998), the mean value of $\rm V_{\rm 
Val}$ is given for
comparison, $t_{\rm int}$ is the integration time in drift scanning mode and N is 
the number of processed images
for each year. 

From the original images, fields of approximately 13' x 13' (512 x 512 pixel) were 
extracted. 
Data reduction was made with the IRAF package\footnote{IRAF is distributed by the 
National Optical 
Observatories which is operated by the Association of Universities for research in 
Astronomy, Inc. 
under co-operative agreement with the National Science Foundation},
 using the APPHOT task for aperture photometry in the usual mode, with a median 
model
to discount the sky contribution and an aperture of three to four times the FWHM. 
The chosen aperture
followed the recommendations of Cellone et al. (2000), to avoid the effects in the 
resulting
light curves of the host galaxies, whose contribution can change due to seeing 
fluctuations. 

Since these light curves were constructed in differential mode, the stability in 
magnitude of 
the comparison
 and control stars is an important factor. Our database with
a large time coverage allowed us to quantify this stability.
The control stars are the set of field stars chosen to construct a stable 
reference light
curve, represented by the mean magnitude. The stability of the comparison star,
from which the differential light curve is constructed, was obtained comparing its
magnitude to the mean magnitude of the control stars. 
Table 2 presents information concerning these stars, most of them are from the 
Tycho II catalog (H\o g et al. 2000). 
In the first column are the names of
the stars and in the second and third
their coordinates, the astrometric precision is better than 0.05 arcseconds in 
average.
 The fourth column gives the Valinhos magnitude and
  the last column shows the standard deviation
weighted with the individual errors, which are not dependent on the real magnitude 
values
and are good quantitative measures of magnitude 
stability.

 The data selected to construct the light curves for variability analysis 
have a confidence level
 of more than $2\sigma$, where $\sigma$ is the dispersion of the control light 
curve. 
However, the errors presented in the light curves were obtained directly from the 
aperture 
photometry and, in average, are larger than $\sigma$.
A few data points in the  light curve of  PKS 2005-489 seem to follow the same 
pattern as
 the control light curve, but were not eliminated because they obey the $2\sigma$ 
criteria.

For the photometric nights, the instrumental magnitudes  were calibrated, 
transformed 
to the Johnson system  and then to flux densities, after correcting by 
interstellar reddening
using the values for visual absorption $A_{V}$ derived by Bersanelli et al. 
(1992).
The calibration was based on Tycho II stars (H\o g et al. 2000)
present in the observed fields of the 
blazars, using the standard procedure adopted in Valinhos's meridian circle data 
analysis,
as described in Dominici et al. (1999) and Viateau et al. (1999).

\section{The resulting light curves}

Figure 1 shows the light curve of PKS 2005-489 covering all the campaigns.
Brightness variations were observed at several time scales, ranging from a few 
days to years.
In particular, even though the time coverage was irregular, 
we could identify in several occasions similar flare-like variations with monthly 
time scales. 
 Figure 2 shows five possible 
 peaks superposed and arbitrarily shifted in magnitude and time to
emphasize their apparent similarity in shape and duration. 

The largest variation in the light curve (about one magnitude in 20 days)
was observed in 1998, close to the occurrence of the strong X-ray flares. 
Except for this short time variation,
the mean flux density was lower than those measured in the others years.
The behavior of the source just before the X-ray flare that started at the 
beginning of September 2000 
(our observations ended in August 30) was characterized by an increase in optical 
brightness,
also observed with better time resolution by Rector \& Perlman (2003). 
Moreover, the mean flux density for that year was 0.5 magnitude brighter than in 
1998.

The light curve of PKS 2155-304 between 1996 and 1999 is shown in Figure 3.
It covers some time before and after the TeV and X-ray flares of
November 1997 (Chadwick et al. 1999). Although  our time coverage is limited,
there seems to be a dimming from 1996 to 1998, followed by a chaotic behavior in 
1999, when very
rapid variations were observed. This is not surprising, since 
PKS 2155-304 is known to be an intraday variable
source (Paltani et al. 1997, for example). 
However, the mean flux density did not show significant changes in the four years 
of observation. 
The maximum variation in our observed 1996 light curve was 0.25 magnitudes, in 
agreement with the 0.35
magnitudes reported by Bertone et al. (2000) as 
the results of a short duration multiwavelength campaign in May 1996, 
just before the beginning of our observations.

We must point out the
presence of at least two very sharp dips in the light curve of 1999, with duration 
of about two days, that are shown in detail in Figure 4.  
Other dips were reported in the literature at  several  wavelengths in 1994
(Pesce et al. 1997, Urry et al. 1997)
and in polarized light in 1994 and 1998 (Urry et al. 1997, Tommasi et al. 2001).

The mean flux densities at $\nu = 6.5 \times 10^{15}$ Hz, considering all 
campaigns, 
 were 30.8 and 23.0 mJy for
PKS 2005-489 and PKS 2155-304 respectively. These values lie between the maximum 
and minimum
known flux densities for both sources, as collected by Xie et al. (1998). 
For PKS 2005-489, the mean flux density measured in 1998 coincides with that 
predicted by the  SED
modeled by Padovani et al. (2001), using contemporaneous infrared and X-ray data.
However, in 1996 our observed flux density was larger than the value predicted by 
the same authors,
based only in X-ray data.

\section{Correlation with X-ray light curves}

Correlation of our light curves with data at other
frequencies could open more possibilities to the interpretation of the detected 
brightness variations.
For that purpose, we compared our optical observations with the daily averaged 
X-ray data set in the 
1.5 - 12 keV range from RXTE\footnote{http://xte.mit.edu} in the All-Sky Monitor 
project
(ASM, Levine et al. 1996).  As 
X-ray observations present large error bars and our time coverage is in general 
more sparse, 
we averaged the X-ray data at several time intervals,
 calculating their mean values weighted by their errors. 
 The re-sampled X-rays  were then compared with the optical light curves to search 
for correlations at
 different time scales, 
 using the Discrete Correlation Function (DCF), adequate for unevenly sampled time 
series. 
Details of the method could be seen in Edelson \& Krolik (1988).

In Figures 5 and 6 we show the optical and
X-ray data for PKS 2005-489 and PKS 2155-304, respectively, and in Figures 7 and 8 
we present the respective
DCF functions including all epochs. We can see that there is no significant 
correlation between 
the light curves at time scale of years, independent of the chosen time bins.

We then investigated correlations in time scales of several days in the years 
for which a reasonable number of optical data points were available.
The DCF functions of PKS 2005-489 for 1997, 1998 and 2000 are presented in Figure 
9. 
In 1997 a significant correlation was
obtained, with optical emission preceding X-rays; the  time lag was about 30 days 
but the width of the
DCF function did not give a  good time resolution. In 1998, on the other hand, the 
X-rays preceded the 
optical variability
by approximately ten days and  no significant correlation was found between the 
light curves in 2000. Although looking at the optical and X-ray light curves for 
the individual years  one could believe that the significance in the correlations 
is due to a few high intensity points, this does not seem to be the case, as 
verified in the DCF function, which remained almost unchanged when  those 
particular data points were excluded from the calculation.

In the case of PKS 2155-304, only for 1999 there were   sufficient optical data 
points to carry out 
the correlation
analysis, which is presented in Figure 10.  The results indicate a significant  
correlation but it is not 
possible to determine an eventual delay due to the width of the DCF function. 

Assuming that the X-rays are produced by the synchrotron process, variations 
preceding the optical 
counterpart 
can be explained by several models, as time-dependent
inhomogeneous accelerating jets  (Geoganopoulos \& Marscher 1998) or  shocks 
propagating along the jets 
(Blandford \& K\" onigl 1979,
Marscher 1987). 
The same models can explain the situation  in which the optical counterpart 
appears first, if most of
the X-rays were produced by the inverse Compton process, involving synchrotron 
radio or infrared photons.

\section{Dips in the light curve of PKS 2155-304}

As mentioned in section 3, we observed several dips in the 1999 light curve of PKS 
2155-304, and others were
reported previously in the literature. Although a fast decrease in emission can be  
explained by rapid energy loss 
from the relativistic electrons, its fast subsequent rise to the previous flux 
density level is more difficult 
to understand.
For this reason we investigated if eclipses  can be the cause of the observed 
dips. If the
eclipsing source is a broad line region cloud, with size 10$^{16}$ cm and velocity 
of 3000 km s$^{-1}$, 
entering the line of sight to the blazar, the time scale for occultation will be  
of the
order of a year, too large to explain our light curve. On the other hand, if it is 
the eclipsed region
that is moving, as in the case of a shock wave propagation along the relativistic 
jet, with apparent
velocity of the order of $c$, the transit time behind the cloud will be of the 
order of a few days, as
observed. Recent VLBI observations (Piner \& Edwards, 2004) showed that a new 
superluminal feature was formed in 1999,
which could be our eclipsed source.

The differences in the dip strength at different wavelengths (Pesce et al. 1997, 
Urry et al. 1997) could be explained if the
eclipsed region and the quiescent jet have different spectral indices. 
This scenario explains  also the polarization effects observed
by  Urry et al. (1997) and Tommasi et al. (2001). As it is well known from radio 
observations
with high spatial resolution, the polarization properties 
of the synchrotron radiation in AGNs vary along the jet, especially across the 
shocks, where the direction
and intensity of the magnetic field change drastically. Low resolution 
observations provide only the average value
of the Stokes parameters, and therefore, when one part of the emitting region is 
eclipsed, the observed average will
change with the same time scale.

\section{Implications for the SED}

The spectral energy distribution (SED) in BL Lacs is usually attributed to 
emission from a beamed
relativistic jet, oriented in a direction very close to the line of sight (Urry \& 
Padovani 1995). 
The shape of the SED, in the $ {\rm log}\, \nu F_{\nu} \times  {\rm log}\, \nu$ 
representation, 
usually shows two broad 
components: the first, with the peak at lower energies, is interpreted as  
synchrotron emission
 and the second  as inverse Compton  by either the photons originated in the 
synchrotron 
process (Synchrotron Self Compton, SSC) or by external photons (External Compton, 
EC).
The main problem to model the SED is the lack of simultaneous (or at least 
contemporaneous) flux density measurements in a large range of energies.

Fortunately, the two studied BL Lacs were observed in contemporaneous epochs by 
BeppoSAX (Boella et
al. 1997), allowing us to construct  their spectral energy distribution. 
PKS 2005-489 was observed 
in September 1996 and September-December 1998
(Padovani et al. 2001, Tagliaferri et al. 2001) and PKS 2155-304
in November 1996, 1997 and 1999 (Giommi et al. 1998, Chiappetti et al. 1999, Zhang 
et al. 1999).

Instead of using directly the SED to analyze variability, we started with the flux 
density
distribution, as presented in Figures 11a and 12a for  PKS 2005-489 and 
 PKS 2155-304, respectively.
Although the observation are only contemporaneous, it seems that the  relative
amplitude of the optical variability, if any, is much smaller than that of  
X-rays. 
We found that in all cases  a smooth curve can describe the flux
density distributions, implying in a gradual variation of the spectral index
from radio to X-rays. For that reason,
a parabolic fit was made for each epoch  in the 
${\rm log} \,F_{\nu} \times {\rm log} \, \nu$ plane. This procedure has the 
advantage of being independent of
assumptions about the electron energy distribution, which is generally modeled 
with constant spectral indices
and breaks at arbitrary electron energies.
 
No contemporaneous radio and
infrared observations were available, but data from NED (NASA Extragalactic 
Database) were collected.
For PKS 2005-489 we used the infrared data to constrain the fitting
while for PKS 2155-304 we  used only our optical and BeppoSax data. 
The fittings can be seen as continuous  lines in Figures 11a and 12a for PKS 
2005-489 and PKS 2155-304 
respectively. 

Using this simple flux density distribution we calculated the SED, presented in 
Figures 11b and
12b for PKS 2005-489 and PKS 2155-304 respectively.
The parameters of the fitting in the ${\rm log} \,F_{\nu} \times {\rm log} \, \nu$ 
plane, together
with the frequency of the maximum ($\nu_{max}$) in the SED,  
are presented in Table 3.
For PKS 2155-304, although no infrared data were used in the fit, the obtained SED 
coincided
with the NED IR data at the epoch of quiescent X-ray emission, as can be seen in 
Figure 12c.
As we can see from Figures 11b and 12b, the radio data  lie above the extrapolated 
SED for both objects, implying that 
either the flux density distribution is harder at low frequencies,
 or that part of the radio emission arises from
shocks.
Notice that although most of the radio data obtained in the literature are from 
single dish observations
and therefore could include extended emission, the observed flat spectrum is an 
indication that most of the radio
emission originates in the compact core.

We can see that for PKS 2005-489, the synchrotron peak frequency in the SED  was  
a factor 
of four higher in 1998, when the X-ray flare occurred. For  PKS 2155-304 the peak 
frequency was
a factor of three higher in 1997, 
epoch of  the detected TeV emission and 
X-ray counterpart (Chadwick et al. 1999), but it had its minimum value in 1999, 
epoch of maximum variability in our optical light curve.

\section{Conclusions}

This is the first time that a long term optical light curve for PKS 2005-489 is 
presented. Moreover,
the addition of more optical data to the well studied (and yet poorly understood) 
BL Lac PKS 2155-304 allowed us
to confirm that the two studied objects show high degree of variability
at different time scales.

In PKS 2005-489 we could identify several variability events with duration of 
about 20 days,
although  light curves with better
time sampling are necessary to verify if  in fact they are arising from similar 
phenomena.
 PKS 2155-304 showed a significant activity in short time scales during 1999, with 
amplitude
variations of almost 1 magnitude in  a few days. 
 Of special interest is the presence of
at least two sharp dips in the light curve, with duration of about two days each. 
We interpreted them
as the result of eclipses, caused by BLR clouds or star clusters
in the host galaxy.  To obtain the right time scales we had to assume that the the 
eclipsed region is moving
along the  jet with relativistic velocities. The detection of a superluminal 
component formed
at that epoch gives support to our assumption (Piner \& Edwards 2004).
Although our observations cover some 
months before and after the X and gamma ray activity of 1997, which resulted in 
the detection of
emission at  TeV energies (Chadwick et al. 1999), no unusual behavior in the 
optical light curve
was detected at those epochs.

We searched for correlation between the optical light curves and X-rays  
in the 1.5 - 12 keV range from the All-Sky Monitor project. 
We did not find significant correlation for any of the  sources in time scales of 
years.
 However, in time scales of days, we found in PKS 2005-489 possible correlations 
in the campaigns 
 of 1997 and 1998. In the first year the optical variability preceded 
 the X-rays by approximately 30 days and in 1998 X-rays preceded the optical 
variability by about 10 days. Although
 we believe that the optical emission is always due to the synchrotron process, 
the  X-ray activity
 can be explained  also as synchrotron radiation by several models (Geoganopoulos 
\& Marscher 1998, Blandford \& K\" onigl 1979,
Marscher 1987) only when it precedes the optical variability.
 Otherwise the same models can be applied, but assuming the the X-ray emission is 
due to the inverse Compton
 process involving synchrotron radio or infrared photons. 
 The 1999 optical and X-ray light curves of PKS 2155-304 
seems to be correlated, but it was not possible to determine the presence of time 
delays because 
of the width of the DCF function.

 We also constructed the SED with our optical data and contemporaneous BeppoSAX 
X-ray observations.
First we fitted a simple parabola in the 
${\rm log} \,F_{\nu} \times {\rm log} \, \nu$ plane and then constructed the  SED,
instead of using multi-parametric models, as usually done (eg. Padovani et al. 
2001).
Although the observation are only contemporaneous, it seems that the  relative
amplitude of the optical variability, if any, is much smaller than that of  
X-rays. 
The obtained SEDs confirm that the synchrotron peaks move to higher frequencies 
during the
occurrence of X-ray flares, although the shifts were much smaller than what was 
observed for
other blazars (Pian et al. 1998, Giommi et al. 1999).

\acknowledgments

This work was partially supported by the Brazilian agencies FAPESP, CNPq and 
FINEP. 
We thank the referee for the comments that
helped us to improve significantly this paper and A. Caproni for interesting 
discussions.
 This research made use of the NASA/IPAC Extragalactic Database (NED),
which is operated by the Jet Propulsion Laboratory, California Institute
of Technology, under contract with the National Aeronautics and Space 
Administration.

\clearpage
\onecolumn


\begin{figure*}
\plotone{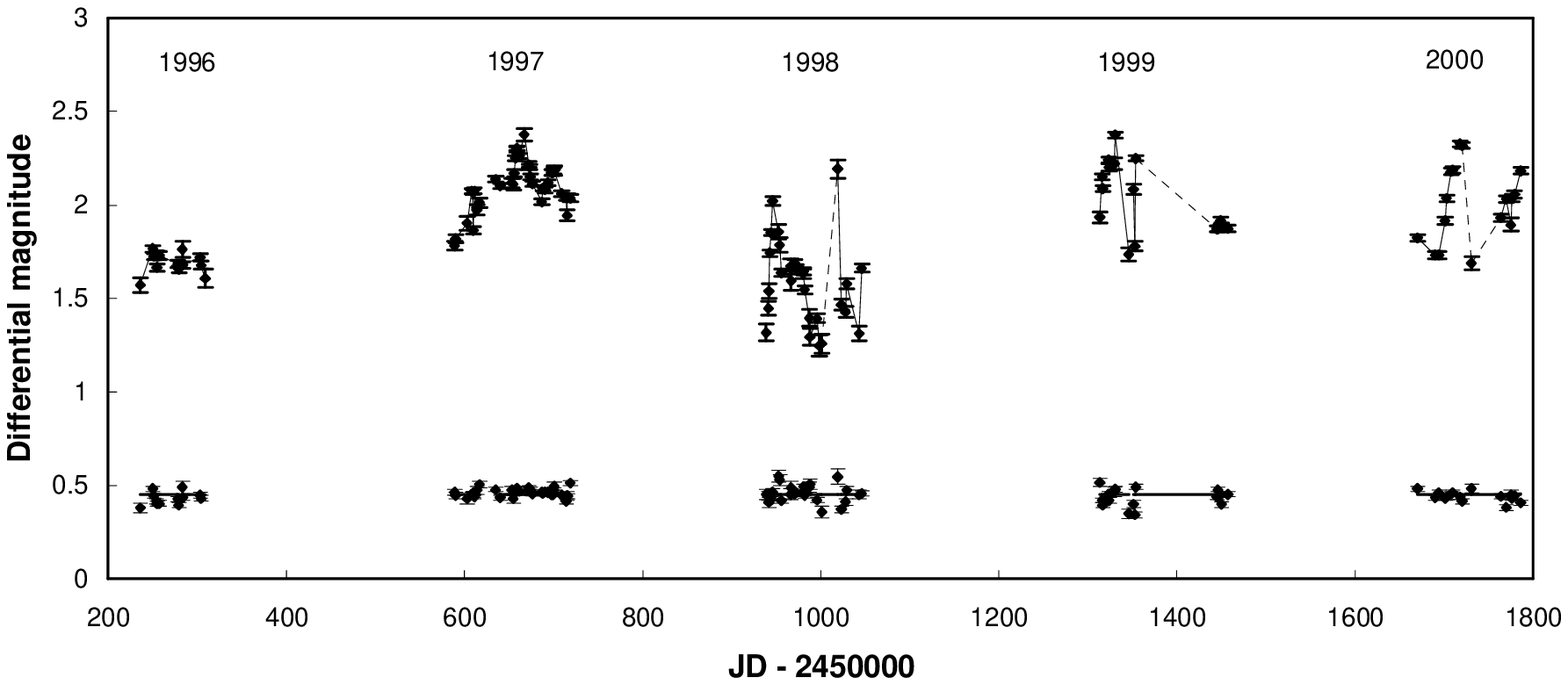}
\caption{Differential light curve of the BL Lac PKS 2005-489 from 
1996 to 2000: upper curve, C2-O, where O is the source. The control light curve
is shown below and was constructed with C2 minus the mean among C1, C3 and C4 (see 
Table 2).
The dashed lines indicate time intervals of more than ten days without 
observations. \label{fig1}}
\end{figure*}

\clearpage 
\twocolumn

\begin{figure}
\plotone{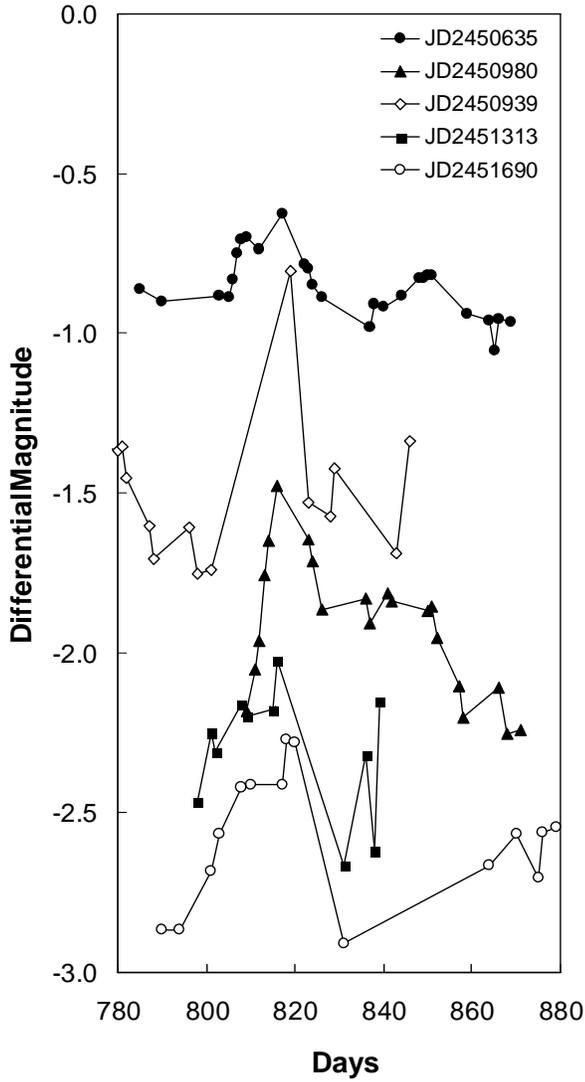}
\caption{Superposition of the peaks detected in 1997, 1998, 1999 and 2000 light 
curves 
of the BL Lac PKS 2005-489. The magnitude scale was arbitrary shifted for better 
comparison.
The legend inside the figure indicated the epoch of the first point in each 
curve.\label{fig2}}
\end{figure}

\clearpage 
\onecolumn

\begin{figure*}
\plotone{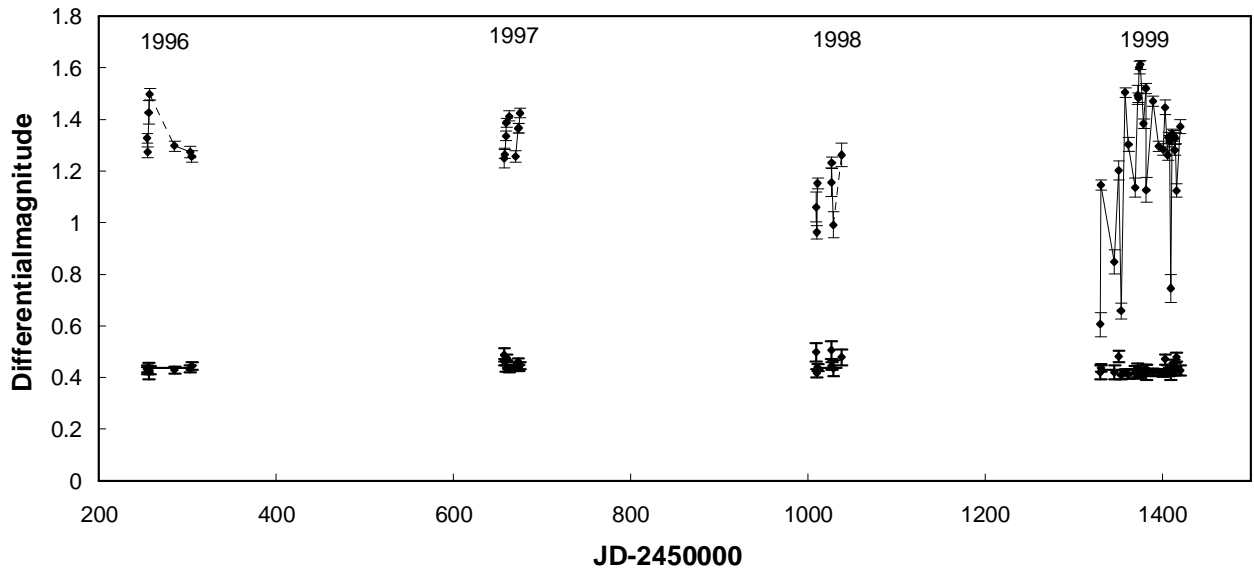}
\caption{Differential light curve of the BL Lac PKS 2155-304 from 
1996 to 2000. Upper curve: C2-O, where O is the source. The control light curve
is shown below and was constructed with C2 minus the mean between C1 and C3 (see 
Table 2).
The dashed lines are as in Figure 1.\label{fig3}}
\end{figure*}

\clearpage 
\twocolumn

\begin{figure}
\plotone{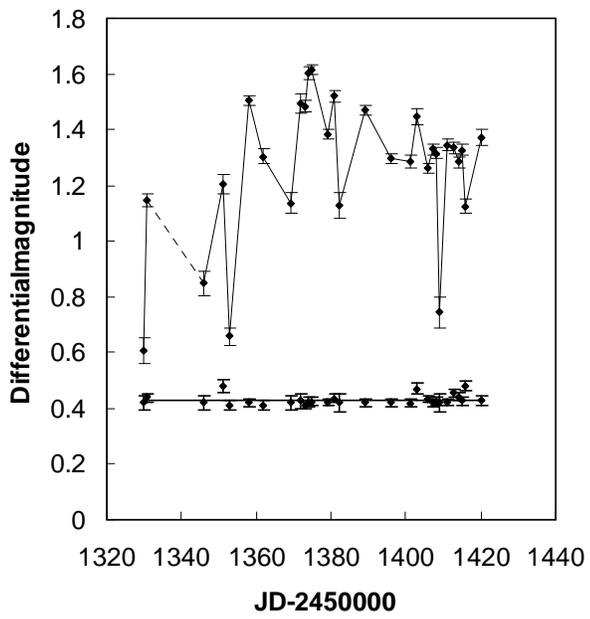}
\caption{Upper curve: differential light curve of the BL Lac PKS 2155-304 in 1999.
Lower curve: the control light curve. The dashed lines are as in Figure 
1.\label{fig4}}
\end{figure}

\clearpage 
\twocolumn

\begin{figure}
\plotone{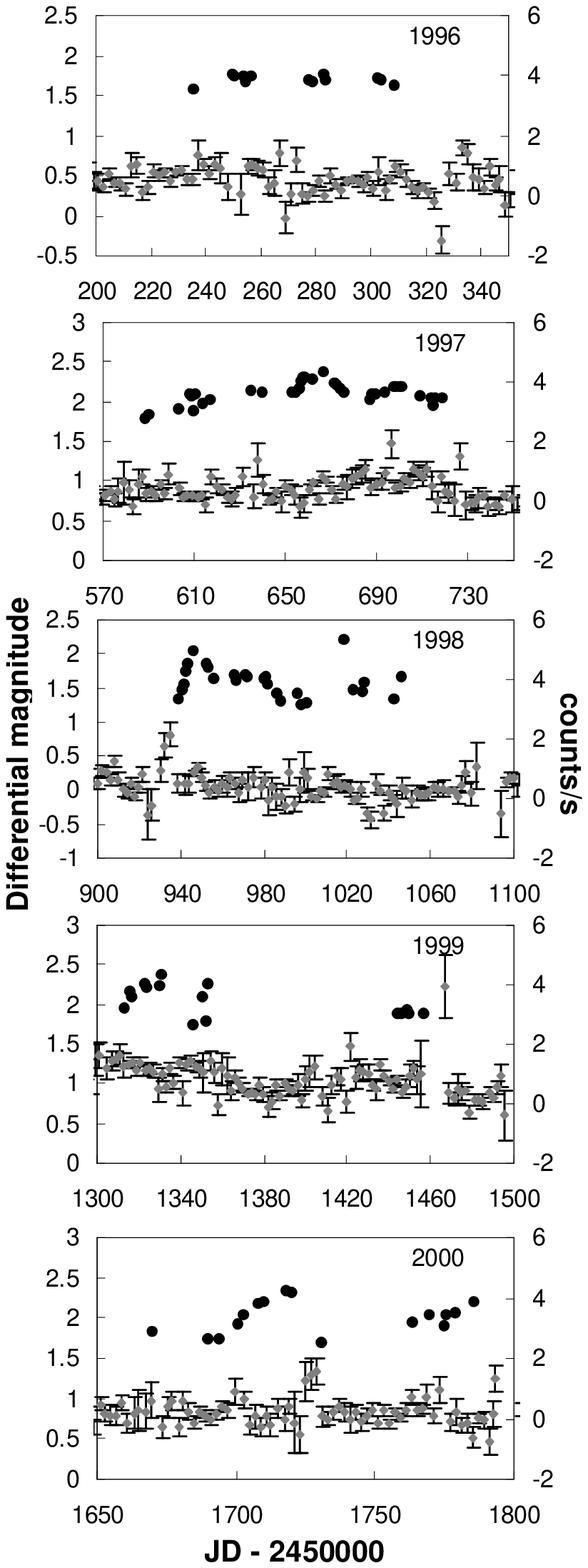}
\caption{Comparison between the optical light curves (black points) of PKS 
2005-489 and the contemporaneous 
X-rays ASM data in the 1.5 - 12 KeV range, from 1996 to 2000. In this example, the 
X-rays data were
binned by calculating the weighted mean value of two points separated by no more
than 4 days\label{fig5}}
\end{figure}

\clearpage 
\twocolumn

\begin{figure}
\plotone{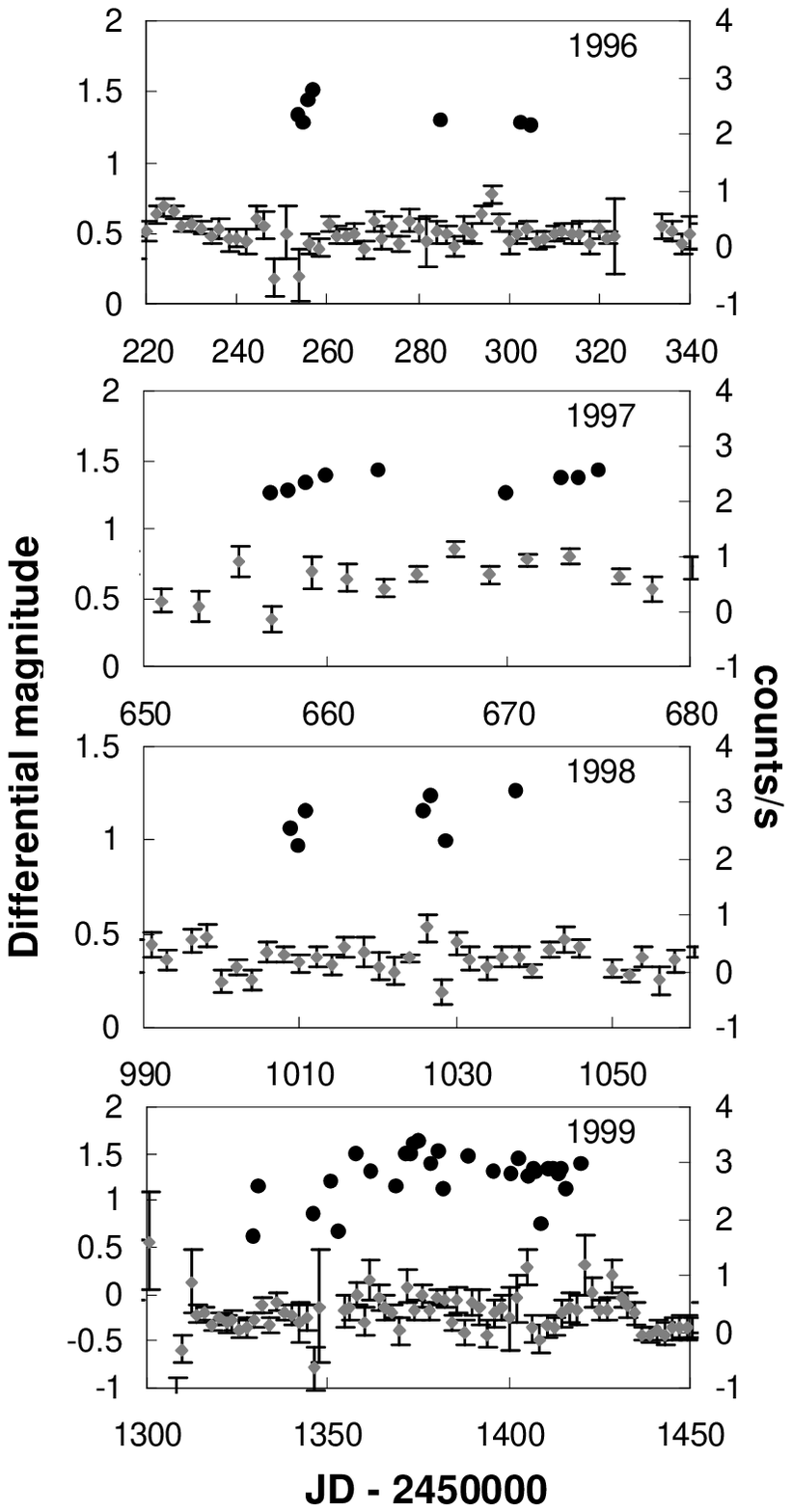}
\caption{Comparison between the optical light curves (black points) of PKS 
2155-304 and the contemporaneous 
X-rays ASM data in the 1.5 - 12 KeV range, from 1996 to 1999. In this example, the 
X-rays data were
binned by calculating the weighted mean value of two points separated by no more
than 4 days\label{fig6}}
\end{figure}

\clearpage 
\twocolumn

\begin{figure}
\plotone{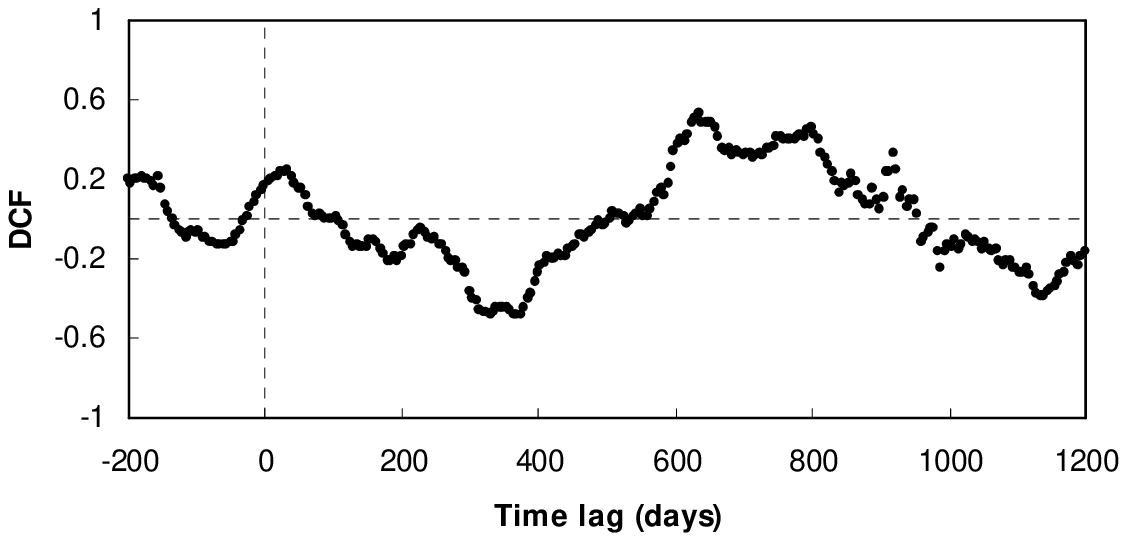}
\caption{Discrete correlation function (DCF) between optical and X-ray data for 
PKS 2005-489 using
the complete dataset and a bin size of 40 days.\label{fig7}}
\end{figure}

\clearpage 
\twocolumn

\begin{figure}
\plotone{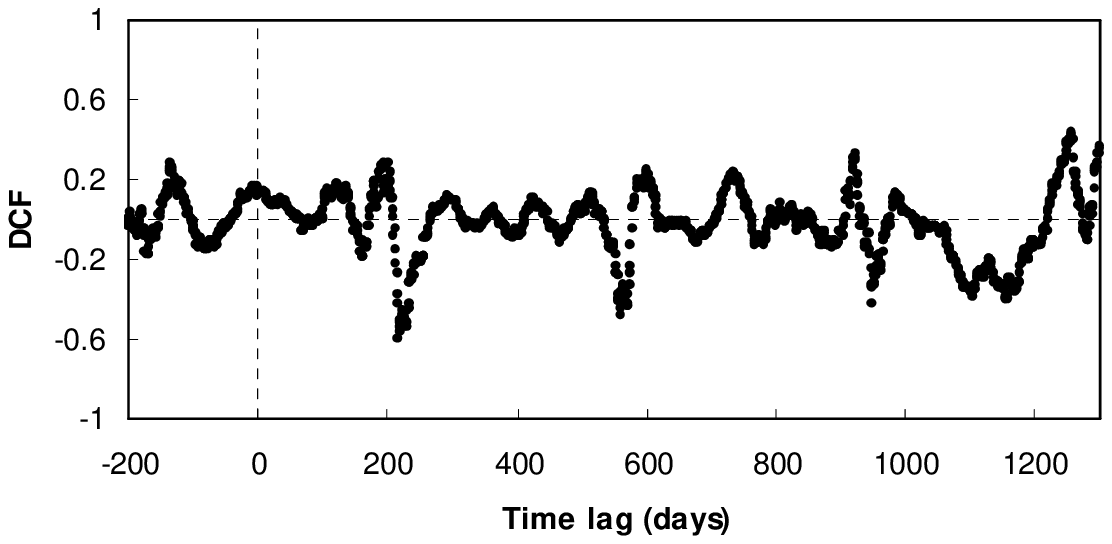}
\caption{Discrete correlation function (DCF) between optical and X-ray data for 
PKS 2155-304 using
the complete dataset and a bin size of 40 days.\label{fig8}}
\end{figure}

\clearpage 
\twocolumn

\begin{figure}
\plotone{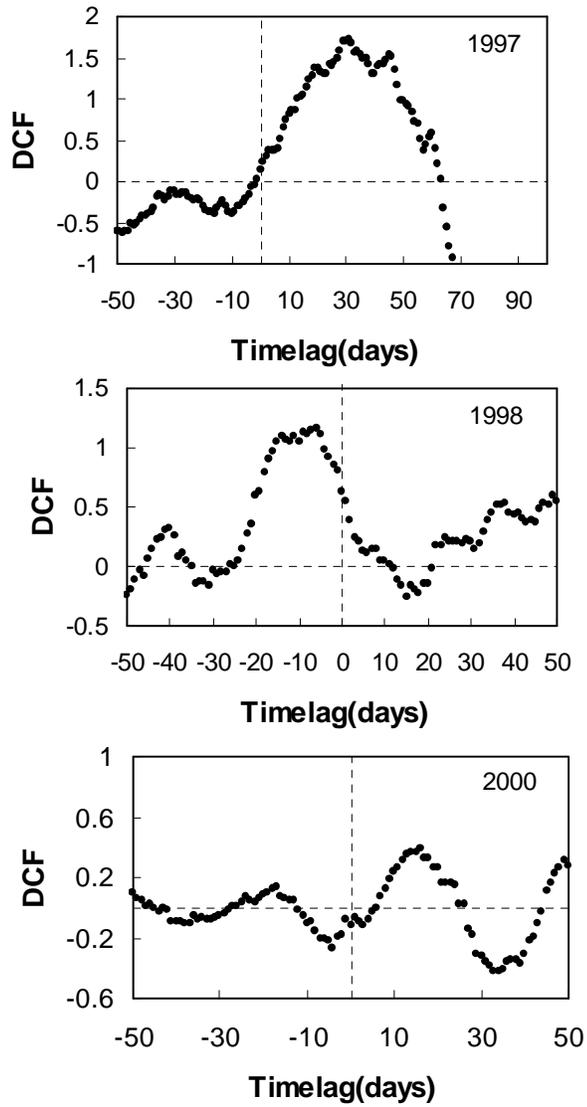}
\caption{Discrete correlation function for PKS 2005-489 in 1997, 1998 and 2000. In 
all cases
the bin size was 20 days.\label{fig9}}
\end{figure}

\clearpage 
\twocolumn

\begin{figure}
\plotone{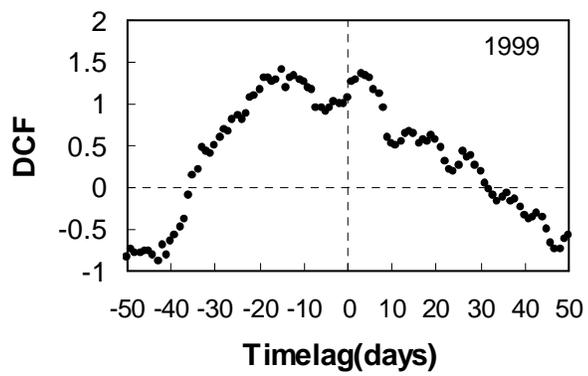}
\caption{Discrete correlation function for PKS 2155-304 in 1999. The bin size was 
20 days.\label{fig10}}
\end{figure}

\clearpage 
\twocolumn

\begin{figure}
\plotone{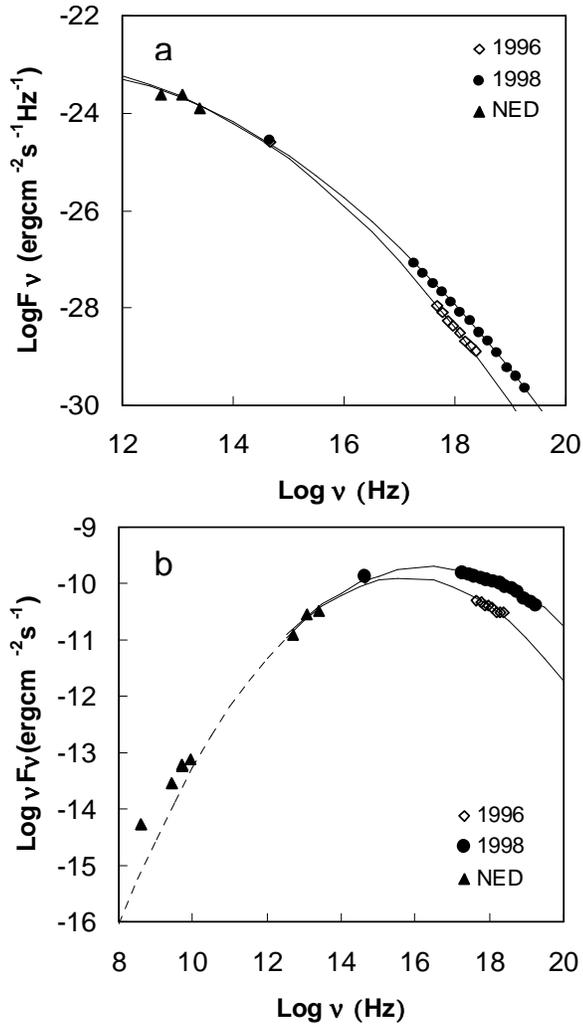}
\caption{Spectral energy distribution of PKS 2005-489 in 1996 and 1998,  using two
different representations:
 a) $ {\rm log}\, F_{\nu} \times  {\rm log}\, \nu$ and b) $ {\rm log}\, \nu 
F_{\nu} \times  {\rm log}\, \nu$. 
The X-ray data are 
BeppoSAX observations presented
by Padovani et al. (2001). The data from infrared and radio wavelengths were
obtained in NED Database. The lines indicate parabolic fits from 
infrared to X-ray data in the $ {\rm log}\, F_{\nu} \times  {\rm log}\, \nu$ plane 
(see text for details).\label{fig11}}
\end{figure}

\clearpage 
\twocolumn

\begin{figure}
\plotone{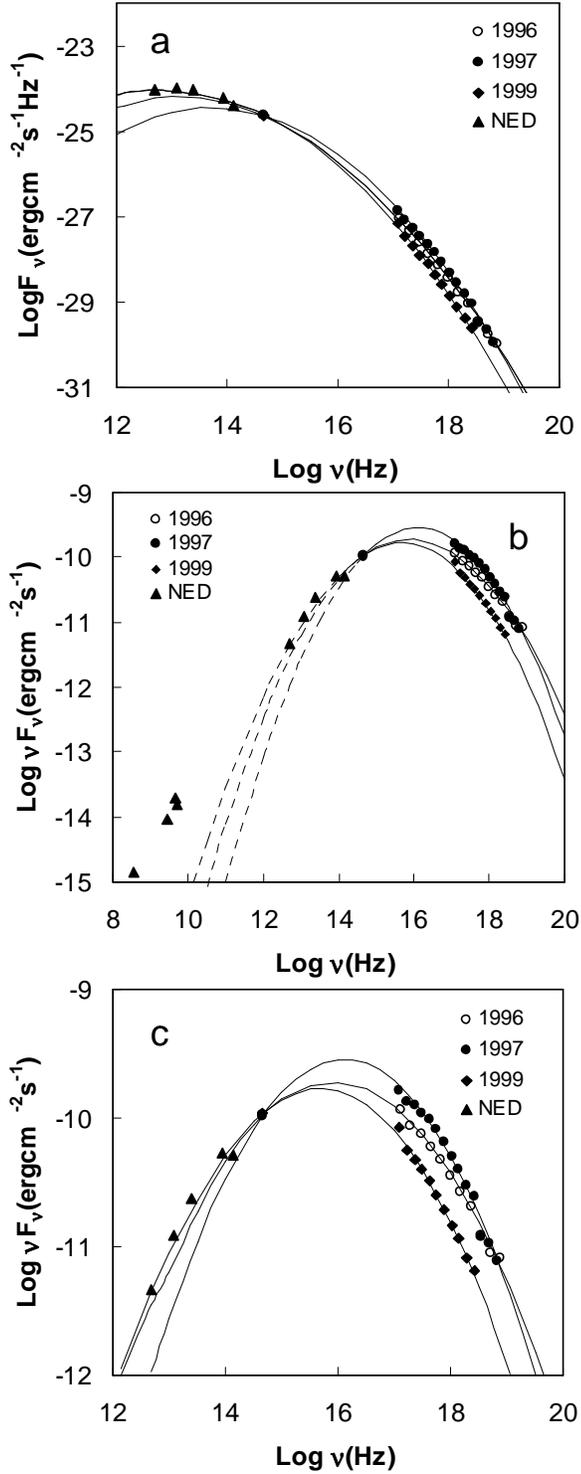}
\caption{Spectral energy distribution of PKS 2155-304 in 1996, 1997
and 1999, using two different representations:
 a) $ {\rm log}\, F_{\nu} \times  {\rm log}\, \nu$ and b) $ {\rm log}\, \nu 
F_{\nu} \times  {\rm log}\, \nu$. 
c) same as b) in an expanded frequency scale.
 The X-ray data are 
BeppoSAX observations presented
by Giommi et al. (1998), Chiappetti et al. (1999) and Zhang et al. (2002). The 
data from infrared and radio 
wavelengths were obtained in NED Database and are not contemporaneous. 
The lines indicate parabolic fits to the
optical and X-ray data in the $ {\rm log}\, F_{\nu} \times  {\rm log}\, \nu$ and 
plane  
(see text for details).\label{fig12}
 }
\end{figure}

\clearpage 
\onecolumn

\clearpage

\begin{deluxetable}{@{}lcrcccccc@{}}
\tabletypesize{\scriptsize}
\tablecaption{Basic data of the observed sources and the observational program.
 \label{tbl-1}}
\tablewidth{0pt}
\tablehead{
\colhead{Object}&  \colhead{$\alpha$  (J2000)} & \colhead{$\delta$ (J2000)} &
\colhead{V}& \colhead{$\rm V_{\rm Val}$} &
\colhead{z} & \colhead{$t_{\rm int}$ (s)} & \colhead{Date}  & \colhead{N}}
\startdata
PKS 2005-489 &    20 09 25.39    &    -48 49 53.7 & 13.4   & 13.0  & 0.071  & 78.5 
& 05/31 - 08/13/1996& 13 \\  
             &                    &               &        &       &        &       
&   05/19 - 09/27/1997  &  37 \\
             &                    &               &        &       &        &       
&   05/05 - 08/20/1998 & 27 \\
             &                    &               &        &       &        &       
&  05/14 - 10/06/1999  & 16\\
             &                    &               &        &       &        &       
&   05/05 - 08/30/2000  & 16  \\
\hline
PKS 2155-304   &    21 58 52.06    &    -30 13 32.1  & 13.1 &  13.1  & 0.170  & 
59.8 &  06/18 - 08/08/1996  & 7 \\
             &                    &               &        &       &        &       
& 05/31 - 08/13/1997   &  9 \\
             &                    &               &        &       &        &       
&  07/14 - 08/12/1998  & 7 \\
             &                    &               &        &       &        &       
&  05/31 - 08/30/1999 &  29 \\
\enddata

\end{deluxetable}

\clearpage

\begin{deluxetable}{@{}lcrcrr@{}}
\tabletypesize{\scriptsize}
\tablecaption{Stars chosen for reference and control in the field of the studied 
objects. 
\label{tbl-2}}
\tablewidth{0pt}
\tablehead{
\colhead{Object}&  \colhead{$\alpha$  (J2000)} &\colhead{$\delta$ (J2000)} & 
\colhead{$V_{\rm Val}$} &
\colhead{sigma}}
\startdata
\colhead{PKS 2005-489}\\
\hline
C1 (8400   1817) &   20 09 23.25    &   -48 52 30.2 &  9.25   &   0.0003   \\  
C2 (8400   900)&    20 09 05.42    &    -48 47 20.9 & 11.88 &   0.0018  \\
C3 (8400   620)&    20 09 48.72    &    -48 48 51.6   & 12.20 & 0.0018 \\
\hline
\colhead{PKS 2155-304}\\
\hline
C1 (7488   124) &    21 59 01.52    &   -30 09 29.9  & 9.18   &   0.0033   \\  
C2 (7488   340)&    21 59 02.51    &    -30 10 46.6  & 12.14 &   0.0044  \\
C3             &    21 59 05.33   &    -30 10 51.1    & 12.96 & 0.0069 \\
C4             &    21 58 46.51    &   -30 17 51.3   & 12.76 & 0.0062 \\
\enddata

\end{deluxetable}

\clearpage

\begin{deluxetable}{@{}lccccccc@{}}
\tabletypesize{\scriptsize}
\tablecaption{Parameters of the parabolic fits and frequency of the synchrotron 
peak
for each epoch \label{tbl-3}}
\tablewidth{0pt}
\tablehead{
\colhead{Object}&  \colhead{Epoch} & \colhead{quadratic} &
\colhead{linear} & \colhead{constant}  &  \colhead{$\chi^{2}$/dof}  & 
\colhead{$\nu_{peak}$ (Hz)}}
\startdata
PKS 2005-489 &1996 &-0.10 & 2.521 & -35.19 & 0.04/10 &$5.6 \times 10^{15}$ \\  
              &1998 &-0.08  & 1.58 & -30.89 & 0.02/15 &$2.5 \times 10^{16}$ \\    
\hline
PKS 2155-304   &1996 &-0.16  & 4.07 & -50.00  & 0.02/9 &$7.9 \times 10^{15}$  \\  
              &1997 & -0.21  & 5.76 & -63.93 & 0.03/15 & $1.3 \times 10^{16}$  \\  
              &1999 & -0.20  & 5.18  & -58.26 & 11.31/9 &$5.0 \times 10^{15}$  \\  
\enddata

\end{deluxetable}

\end{document}